\documentclass[aps,prl,twocolumn,superscriptaddress,nofootinbib,longbibliography,10pt]{revtex4-2}

\usepackage{amsmath}
\usepackage{amsfonts}
\usepackage{enumerate}
\usepackage{amsfonts}
\usepackage[utf8]{inputenc}
\usepackage[T1]{fontenc}

\usepackage{bbm}
\usepackage{amssymb}
\usepackage{amsthm}
\usepackage{mathtools}
\usepackage{comment}
\usepackage{mathrsfs}

\usepackage[dvipsnames]{xcolor}
\definecolor{linkcolor}{rgb}{0.0,0.3,0.5}
\usepackage[unicode, colorlinks=true, linkcolor=linkcolor, citecolor=linkcolor, filecolor=linkcolor,urlcolor=linkcolor, pdfusetitle]{hyperref}

\makeatletter
\def\@fpheader{\relax}
\makeatother


\newcounter{parentsubequation}

\makeatother

\DeclareMathAlphabet{\mathbbold}{U}{bbold}{m}{n} 

\DeclareMathOperator{\Tr}{Tr}

\DeclareMathOperator{\adj}{adj}

\begin{document}

\title{Ergodic Hysteresis of the Kerr black hole spectrum} 

\author{João Paulo Cavalcante}
\email{joao.cavalcante@ufabc.edu.br}

\affiliation{Centro de Matemática, Computação e Cognição,
Universidade Federal do ABC (UFABC), 09210-580, Santo André, São Paulo, Brazil}

\author{Maurício Richartz}
\email{mauricio.richartz@ufabc.edu.br}

\affiliation{Centro de Matemática, Computação e Cognição,
Universidade Federal do ABC (UFABC), 09210-580, Santo André, São Paulo, Brazil}

\author{Bruno Carneiro da Cunha}
\email{bruno.ccunha@ufpe.br}

\affiliation{Departamento de Física, Universidade Federal de
  Pernambuco, 50670-901, Recife, Brazil}

\begin{abstract}
	\noindent We uncover a cascade of exceptional points (EPs) in the
	quasinormal mode spectrum of massive scalar perturbations of Kerr
	black holes, revealing an intricate non-Hermitian structure underlying
	their linear response. The cascade originates from a single damped
	mode that enters the extremal spectrum for sufficiently large field
	masses. We obtain evidence for an infinite sequence of EPs in the
	$(\ell,m)=(1,1)$ and $(2,2)$ sectors near the extremal limit,
	mediating the transition between damped and zero-damping modes. Each
	EP carries a geometric phase that enables adiabatic mode mixing across
	the entire overtone spectrum, a phenomenon we refer to as adiabatic
	ergodicity. 
\end{abstract}
\maketitle

\noindent \textbf{\textit{Introduction --}}  Quasinormal modes (QNMs), and their associated complex frequencies
$\omega$, characterize the damped oscillations through which systems,
ranging from black holes to vortex flows, respond to external
perturbations and lose energy~\cite{Ching:1998mxl, Kokkotas:1999bd,
Nollert:1999ji, Berti:2009kk, Konoplya:2011qq, 2017PhRvX...7b1035A,
Berti:2025hly, Jacquet:2021scv, Torres:2020tzs}. In gravitational
systems, QNMs may be excited, for instance, through test
fields~\cite{Vishveshwara:1970zz,ZOUROS1979139}, infalling
particles~\cite{Davis:1971gg}, and
mergers~\cite{LIGOScientific:2016aoc}. In vacuum General Relativity, QNM frequencies
for each set of mode numbers $(\ell, m, n)$ depend only on the mass
$M$ and the spin $a$ of the black hole, thereby providing the basis
for black hole spectroscopy~\cite{Echeverria:1989hg, Dreyer:2003bv,
Berti:2016lat}. An interesting feature of the QNM spectrum emerges
as the extremal limit ($a \rightarrow M$) is
approached~\cite{Yang:2012pj,Yang:2013uba}. In this regime, the
spectrum branches into zero-damping modes (ZDMs) and damped modes
(DMs). ZDMs are characterized by $\omega \rightarrow m/(2M)$ and a
vanishing imaginary part as $a \rightarrow M$. Unlike DMs, these modes
are long-lived in the extremal limit and play a crucial role in
analyses of black hole
stability~\cite{Yang:2014tla,Gralla:2016sxp,Casals:2016mel,Richartz:2017qep,Casals:2019vdb,TeixeiradaCosta:2019skg,Klainerman:2022ric,Shlapentokh-Rothman:2023bwo}. 

The eigenvalue problem underlying black hole perturbations is
non-Hermitian. Hence, eigenvalue degeneracies that manifest as
exceptional points (EPs) are, in principle,
permitted~\cite{Kato:1995,Ashida02072020}. Such EPs give rise to
remarkable phenomena across different areas of physics, including
spectral sensitivity~\cite{Yang2023Spectral} and geometric
phases~\cite{Cohen:2019hip}. Although the study of EPs in
gravitational physics is still incipient, recent
work~\cite{Davey:2023fin, Dias:2022oqm, Dias:2021yju,
PhysRevLett.134.141401, PhysRevD.110.124064,PhysRevLett.133.261401}
has begun to connect black hole perturbations to the rich EP
phenomenology observed in open quantum systems~\cite{Am-Shallem_2015},
condensed matter~\cite{Liu2019ExceptionalPoints}, and
photonics~\cite{PhysRevA.105.062214}. In particular, the QNMs of a massive scalar field around a near extremal Kerr black hole exhibit EPs and geometrical phases when the parameters are varied adiabatically~\cite{PhysRevD.110.124064,PhysRevLett.133.261401}. The geometrical phase manifests as an interchange of states in the QNM spectrum, producing hysteretic behavior: the ordering of the states depends on the path taken in parameter space. Here, ``adiabatically'' denotes a slow and continuous variation of the parameters, namely the black hole spin and the field mass.

Expanding on these developments, Ref.~\cite{ht2n-vvvh} analyzed EPs
(and avoided crossings) in the Kerr and Kerr–de Sitter spectrum. It
was shown that, while mode amplitudes can be resonantly enhanced near
EPs, modes undergoing avoided crossings interfere destructively,
preserving the stability of the ringdown signal. Ref.~\cite{hfv8-n444}
conducted a numerical study of resonances between the fundamental mode
and the first overtone in a simplified model, providing further
evidence of EP behavior and demonstrating that QNM frequencies cannot
be accurately constrained without accounting for resonance
effects. Ref.~\cite{f8m8-vr4l}, working with a novel method to
determine QNMs of Type-D black holes, confirmed the existence of an EP
between the $n=0$ and $n=1$ modes for massive perturbations of Kerr
black holes, corroborating previous
findings. Ref.~\cite{PhysRevD.111.124002}, on the other hand, used a
two-level effective Hamiltonian to demonstrate that crossings in the
real parts of QNM frequencies coincide with avoided crossings in the
imaginary parts, with resonant effects strongest near an EP. 

In this Letter, we reveal a global structure of EPs in the QNM spectrum of massive scalar perturbations around Kerr black holes. We show that the $(\ell,m)=(1,1)$ and $(2,2)$ sectors exhibit an infinite cascade of EPs, providing a concrete physical realization of such a structure. The geometric phases associated with these EPs induce a global mixing of the overtone spectrum: under suitable closed adiabatic variations of the system parameters, any QNM state can be mapped to any other. By carefully tracking the relevant QNM branches as the parameters are varied, we trace the origin of the entire EP sequence, and its associated geometric phases, to a single DM that enters the extremal spectrum for sufficiently large scalar field mass $\mu$. More broadly, the existence of this infinite EP cascade uncovers a previously unexplored mechanism for manipulating the QNM spectrum during black hole ringdown through suitable evolution in parameter space.

Throughout this work we adopt natural units ($G = c = \hbar = 1$).

\noindent \textbf{\textit{QNM problem via the Isomonodromic method --}} 
Recent developments in the theory of the differential equations governing quasinormal modes have emphasized the key role of the global analytic structure of the solutions. This structure is best framed in terms of the monodromy parameters, which, for second order differential operators with algebraic potentials, can be related to quantum periods of the Seiberg-Witten curve used in certain four-dimensional supersymmetric field theories~\cite{Aminov:2020yma,Bonelli:2021uvf,Consoli:2022eey}. This relation led to connection formulas for the Heun equation and its confluent limits \cite{Bonelli:2022ten,Lisovyy:2022flm}, as well as explicit series solutions for the Jimbo-Miwa-Ueno tau function of isomonodromic deformations ~\cite{ablowitz2006solitons,its2006isomonodromic,Jimbo:1981aa,Jimbo:1981ab,Jimbo:1981ac,Gamayun:2013auu}. The latter has enabled efficient numerical computation of QNMs in a variety of black holes~\cite{Novaes:2014lha,CarneirodaCunha:2015hzd,Novaes:2018fry,CarneirodaCunha:2019tia,Amado:2020zsr,daCunha:2021jkm,Cavalcante:2021scq,Amado:2021erf,daCunha:2022ewy,cavalcante2023isomonodromy}.

In the case of massive scalar perturbations around Kerr black holes, suitable changes of variables reduce both the radial and angular wave equations to the confluent Heun equation (CHE)~\cite{PhysRevLett.133.261401}
\begin{equation}
\frac{d^2y}{dz^2}+\left[\frac{1-\theta_1}{z}+\frac{1-\theta_2}{z-t}\right] \frac{dy}{dz}-
\left[\frac{1}{4}+\frac{\theta_\star}{2z}+\frac{t\,c_t}{z(z-t)}\right]
y=0.
\label{eq:che}
\end{equation}
For the radial equation, the monodromy parameters $\{\theta_1, \theta_2, \theta_\star\}$, together with the accessory parameter $c_t$ and the modulus $t$, are given by~\cite{PhysRevLett.133.261401}
\begin{equation}
\begin{gathered}
	\theta_{1}  = \frac{-i}{2\pi T_{-}}\left(\omega-m\Omega_{-}
	\right), \quad
	\theta_{2}= \frac{i}{2\pi T_{+}}\left(\omega-m\Omega_{+} \right),
	\\
	\hspace{-0.5cm}\theta_{\star}
	=\frac{2iM(2\omega^2-\mu^2)}{\sqrt{\omega^2-\mu^2}}, 
	\quad t=2i (r_+-r_-)\sqrt{\omega^2-\mu^2}, \\ 
	\hspace{-1.2cm} tc_{t}  = \lambda_{\ell,m} + r_+^2\mu^2-(3a^2+r_-^2+3r_+^2)\omega^2 \\ \hspace{2.5cm} + \tfrac{1}{2}t(\theta_2-1) 
	- \tfrac{1}{4}(\theta_{\star} - 2)\theta_{\star},
	\label{eq:parameters}
\end{gathered}
\end{equation}
where $\lambda_{\ell, m}$ is the angular eigenvalue and $r_{\pm}$ denotes the Cauchy ($-$) and event ($+$) horizons, with corresponding temperatures and angular velocities given by $T_\pm=(r_+-r_-)/(8\pi Mr_\pm)$ and $\Omega_\pm=a/(2Mr_\pm)$, respectively.

This framing of the QNM eigenvalue problem allows the exploration of the spectrum across the entire parameter space, covering both the Schwarzschild and extremal limits, as is described in detail in~\cite{PhysRevD.110.124064}. The relevant Julia scripts are made publicly available in~\cite{cavalcante_2024_13961216}. Concretely, the isomonodromic method solves the Riemann-Hilbert map, which relates the monodromy properties of solutions of the generic CHE equation \eqref{eq:che} to its parameters. The non-trivial monodromy data consists of two complex parameters $\sigma,\eta$, and the map is achieved by simultaneously solving two transcendental equations~\cite{daCunha:2021jkm,cavalcante2023isomonodromy}
\begin{subequations}
\begin{gather}
	\hspace*{3.5cm}\tau_V(\theta_1,\theta_2,\theta_\star;\sigma,\eta;t)=0,  \label{eq:rhmap:a}\\
	\frac{d}{dt}\log\tau_V(\theta_1,\theta_2-1,\theta_\star+1;\sigma-1,\eta;t)- \frac{\theta_1(\theta_{2}-1)}{2t}
	=c_t, \label{eq:rhmap:b}
\end{gather}
\label{eq:rhmap}
\end{subequations}
where $\tau_{\tiny V}$ is the tau function for the Painlevé V transcendent. General expansions of $\tau_{\tiny V}$ for small $t$ have been given by \cite{Gamayun:2013auu,Lisovyy:2018mnj} in terms of Nekrasov functions. A Fredholm determinant formulation was derived in \cite{Lisovyy:2018mnj} and adapted to QNM problems in \cite{CarneirodaCunha:2019tia}. In the isomonodromic method, the QNM boundary conditions translate into the following relation between the parameters $\sigma$ and  $\eta$ (valid for $\mathrm{Re}\,(M\omega) >m/2$),
\begin{multline}
e^{\pi i(\eta+\sigma)}=
\frac{\sin\frac{\pi}{2}(\theta_\star+\sigma)}{\sin\frac{\pi}{2}(\theta_\star-\sigma)}\prod_{\epsilon=\pm 1}
\frac{\sin\frac{\pi}{2}(\theta_2+\epsilon \theta_1+\sigma)}{
	\sin\frac{\pi}{2}(\theta_2+\epsilon\theta_1-\sigma)}.
\label{eq:quantcond}
\end{multline}

We remark that, when the black hole is extremal, the radial wave equation for massive scalar perturbations reduces to the double-confluent Heun equation (DCHE) rather than the CHE. The Riemann-Hilbert map \eqref{eq:rhmap} for the DCHE is expressed in terms of the Painlevé III tau function~\cite{Cavalcante:2021scq}. Details relevant to the present discussion are provided in the Supplemental Material~\cite{supplemental}.  

\begin{table}[htb!]
\begin{ruledtabular}
	\begin{tabular}{cccccc}
		$n$ & $(M\mu)^c_n$ & $(a/M)^c_n$ & $n$ & $(M\mu)^c_n$ & $(a/M)^c_n$ \\
		\hline
		0 &  0.37049813 & 0.99946597 & 7   & 0.23336824 & 0.99999142 \\
		1 &  0.31913513 & 0.99985379 & 8   & 0.22797842 & 0.99999328 \\
		2 &  0.29034662 & 0.99993511 & 9   & 0.22342370 & 0.99999460 \\
		3 &  0.27150936 & 0.99996399 & 10 & 0.21951655 & 0.99999556 \\
		4 &  0.25805197 & 0.99997729 & 11 & 0.21612255 & 0.99999629 \\
		5 &  0.24787479 & 0.99998443 & 12 & 0.21314391 & 0.99999686 \\
		6 &  0.23986499 & 0.99998869 & 13 & 0.21050463 & 0.99999730 \\
	\end{tabular}
\end{ruledtabular}
\caption{The first 14 EPs for $(\ell, m) = (1,1)$ massive scalar perturbations of Kerr black holes. The critical values $(M\mu)^c_n$ and $(a/M)^c_n$ are determined by numerically solving Eq.~\eqref{eq:rhmap}. The associated QNM frequencies, which are doubly degenerate, are shown in Fig.~\ref{fig:ImReomega}.}
\label{tab:l1m1_critical} 
\end{table}

\noindent \textbf{\textit{Numerical Results and EP Cascade --}} In previous work \cite{PhysRevD.110.124064,PhysRevLett.133.261401}, we determined the existence of an EP in the QNM spectrum of massive scalar perturbations around near extremal Kerr black holes, which characterizes an accidental degeneracy between the fundamental mode and the first overtone. This EP also signals a transition where the fundamental mode changes from ZDM to DM as $M\mu$ increases. A natural question that follows is how unique this EP is and how general such transitions are. 

By exploring the parameter space in greater detail, we find a series of EPs that sheds light on these issues. Each EP is associated with the degeneracy of a pair of two subsequent overtones $(n,n+1)$ and has a corresponding geometric phase. We label each EP by the index $n$ of the lowest associated overtone and denote the corresponding critical parameters by the superscript $c$, namely $(M\mu)^c_n$ and $(a/M)^c_n$.
In Table \ref{tab:l1m1_critical}, we exhibit the first 14 EPs for $(\ell,m)=(1,1)$, which mix the $(n,n+1)$ pair of overtones up to $n=13$.

\begin{figure}[htb!]
\begin{center}
	\includegraphics[width=0.48\textwidth]{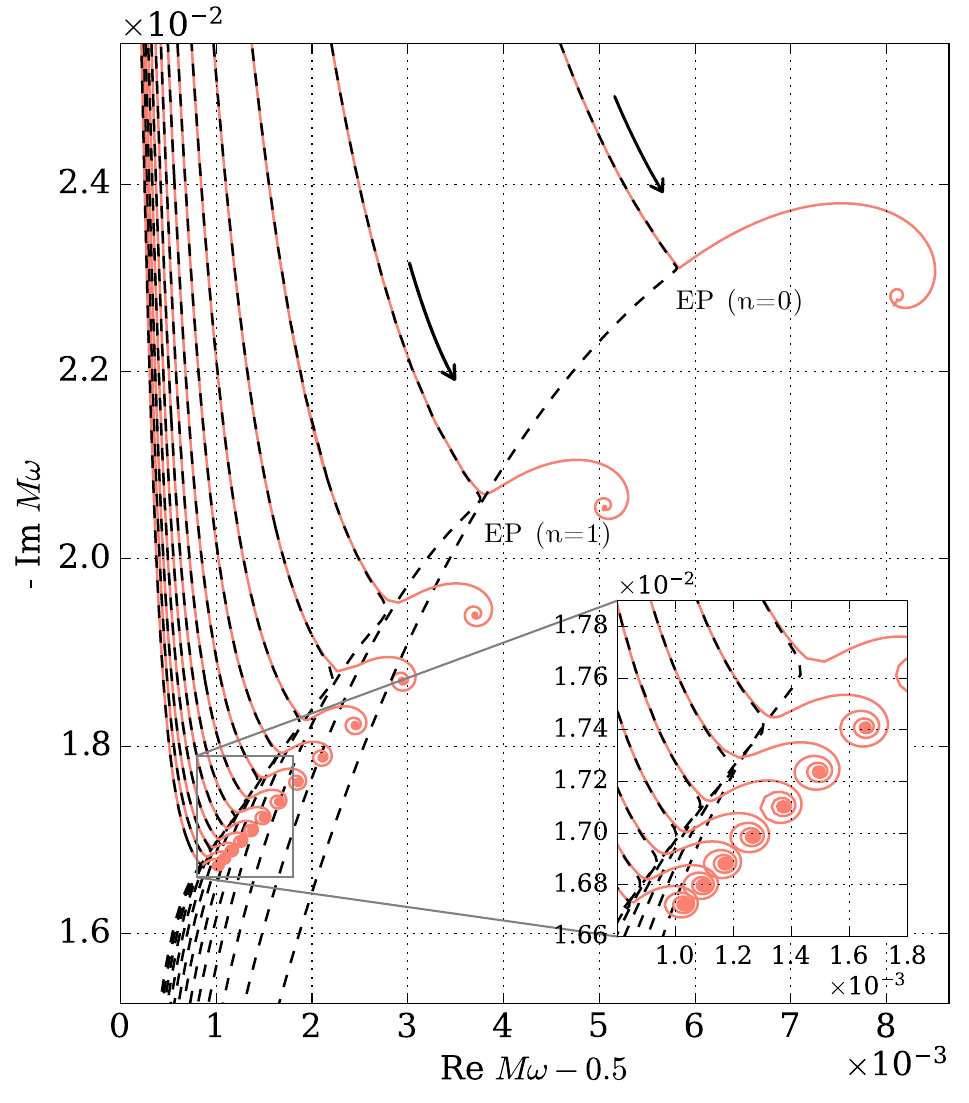}
\end{center}
\caption{
	The behavior of the $(\ell, m) = (1,1)$ overtones, parametrized by $a/M$, near the associated EPs. As the black hole spin increases (along the direction of the arrow), the EP is reached and the initially coincident curves branch out. Dashed curves correspond to scalar masses slightly below the EP value, $M\mu \lesssim (M\mu)^c_n$, and describe ZDMs, which approach $\mathrm{Im}\, (M\omega)=0$ as $a/M\rightarrow 1$. Solid curves correspond to scalar masses slightly above the EP value, $M\mu \gtrsim (M\mu)^c_n$, and describe DMs, which spiral to a finite decay time in the extremal limit. Note that the two masses differ only infinitesimally across the EP, and the corresponding frequencies are nearly degenerate up to the branching point.}
\label{fig:ImReomega} 
\end{figure}

\begin{figure}[htb!]
\begin{center}
	\includegraphics[width=0.45\textwidth]{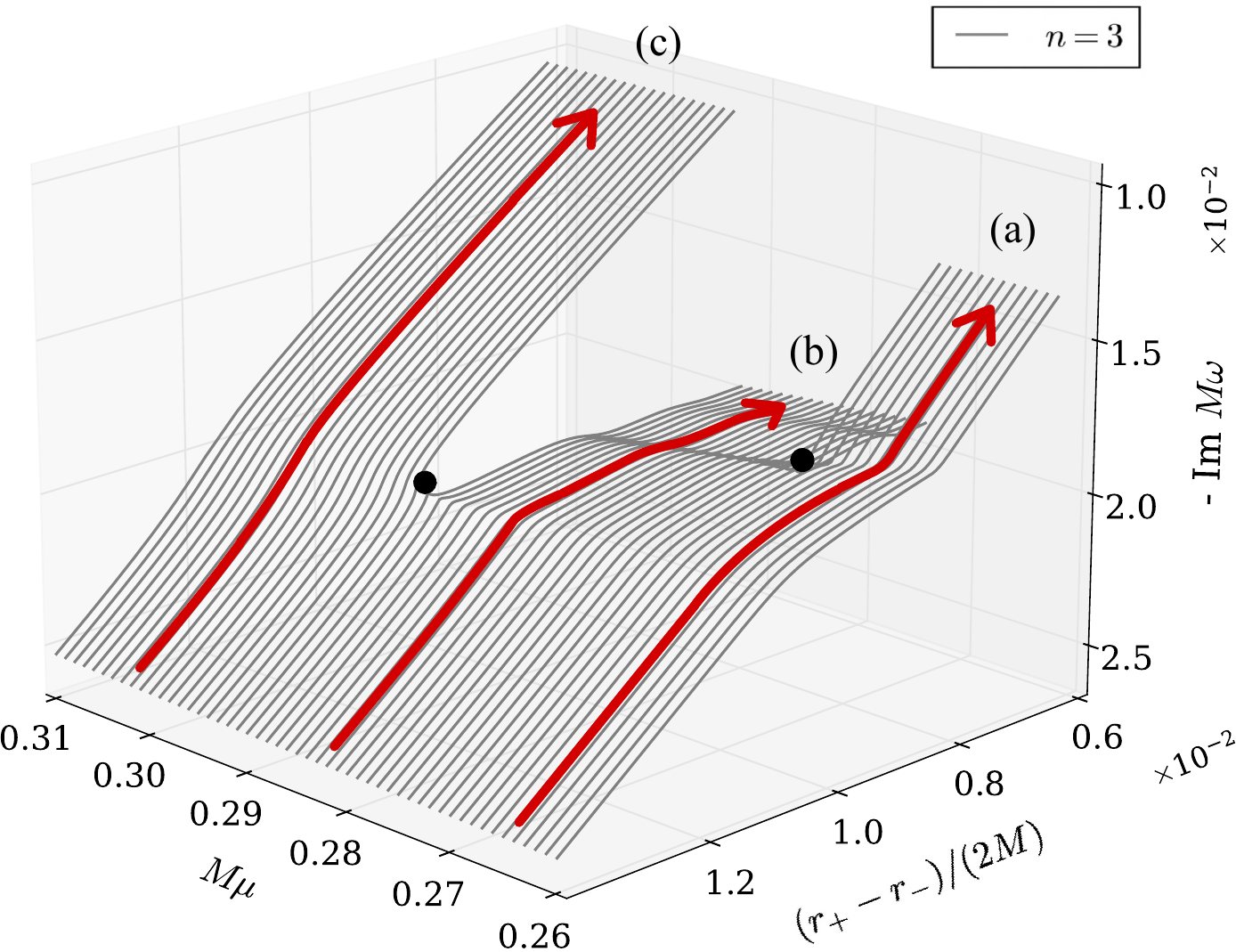}
\end{center}
\caption{The near-extremal behavior of the $n=3$ QNM across a range of $M\mu$ encompassing the $n=2$ and $n=3$ EPs (denoted by dots). The mode transitions from ZDM to DM at each EP. Within region (b), corresponding to the window $(M\mu)^c_3<(M\mu)<(M\mu)^c_2$, the $n=3$ mode exhibits DM behavior, whereas in regions (a) and (c) it behaves as a ZDM.} 
\label{fig:surfacebranching}
\end{figure}

To each EP there is a corresponding ZDM to DM transition, and a parametric plot of the various transitions can be seen in Fig.~\ref{fig:ImReomega}. The EPs form what we refer to as a \textit{degeneracy cascade}. The solid (red) lines, which spiral towards the extremal limit, correspond to DMs. On the other hand, the ZDM behavior is represented by the dashed (black) lines, which can be seen to converge to the asymptotic value $M\omega = m/2=1/2$ as the extremal limit is approached. Additionally, given the associated geometric phases, the two overtones related to a given EP can be transformed into each other if the system parameters are adiabatically varied along a sufficiently small curve around the EP, generalizing the hysteresis phenomenon first revealed in \cite{PhysRevLett.133.261401}. Note that each overtone $n$ may become degenerate with respect to its two neighboring modes ($n\pm 1$) and is, therefore, directly associated with two EPs. Since each EP marks a ZDM to DM transition, this implies the existence of a window $(M\mu)^c_{n+1}<M\mu<(M\mu)^c_n$ in which the $n$-th overtone is a DM for $a/M \rightarrow 1$ (i.e.~$r_+ \rightarrow r_-$), as illustrated in Fig.~\ref{fig:surfacebranching} for $n=3$.

An analytical description of the QNMs around an EP can be obtained from the transcendental equations~\eqref{eq:rhmap}, since the QNM problem is equivalent to solving a non-linear secular equation $\det(\mathbbold{1}-\mathsf{K}(\omega))=0$~\cite{PhysRevD.110.124064}. The integral operator $\mathsf{K}$ is given explicitly in \cite{CarneirodaCunha:2019tia}. Each EP, characterized by the degeneracy of two QNMs, corresponds to a double root of the secular equation. Assuming analiticity of the operator $\mathsf{K}$ with respect to $a/M$ and $M\mu$, the eigenvalue difference near the $n$-th EP is given, to lowest order, by (see the Supplemental Material for details~\cite{supplemental})
\begin{multline}
((M\omega)_{n+1}-(M\omega)_{n})^2\approx \\
A_n((a/M)-(a/M)^c_n)+B_n((M\mu)-(M\mu)^c_n),
\label{eq:nearEPbehavior}
\end{multline}
where $A_n$ and $B_n$ are expansion coefficients. Table~\ref{tab:AnBnl1m1} lists the numerical values of $A_n$ and $B_n$ for the first six EPs, obtained by fitting the above expression to QNM data. These results extend and confirm the findings of Ref.~\cite{hfv8-n444} on the complex square-root behavior near an EP. We point out, as discussed in the Supplemental Material~\cite{supplemental}, that $A_n$ has roughly the behavior of an harmonic series in $n$, evidence for an infinite number of EPs. 

\noindent \textbf{\textit{QNMs: from Schwarzschild to extremal Kerr -- }} At any fixed point in the parameter space $(a/M, M\mu)$, the ordering of the QNM spectrum is unambiguous. One first determines the spectrum by solving the radial and angular equations together, and then orders the modes by the magnitude of the imaginary part of their frequencies. However, the existence of EPs and the associated geometric phases makes such an ordering considerably less robust. As shown in \cite{PhysRevLett.133.261401}, different adiabatic paths can yield different final states, and a single turn around the EP interchanges adiabatically the two levels involved in the accidental degeneracy. The cascade consisting of multiple (possibly infinitely many) EPs allows a complete mixture of the spectrum, as we explain below.   

\begin{table}[htb!]
\begin{ruledtabular}
	\begin{tabular}{ccc}
		$n$ & $A_n \ \times \ 10^{-3}$ & $B_n$  \\
		\hline
		0 & 1.190202 - 0.800304i & 0.104888 - 0.393782i \\
		1 & 0.502592 - 0.416007i & 0.139330 - 0.723933i \\
		2 & 0.289935 - 0.264467i & 0.160232 - 1.047625i \\
		3 & 0.194913 - 0.188481i & 0.175178 - 1.368997i \\
		4 & 0.143415 - 0.144198i & 0.186790 - 1.689281i \\
		5 & 0.111897 - 0.115722i & 0.196266 - 2.008970i \\
	\end{tabular}
\end{ruledtabular}
\caption{Complex expansion coefficients $A_n$ and $B_n$, as defined in \eqref{eq:nearEPbehavior}, associated with the first six EPs.}
\label{tab:AnBnl1m1} 
\end{table}

\begin{figure*}[htb!]
\begin{center}
	\includegraphics[width=0.99\textwidth]{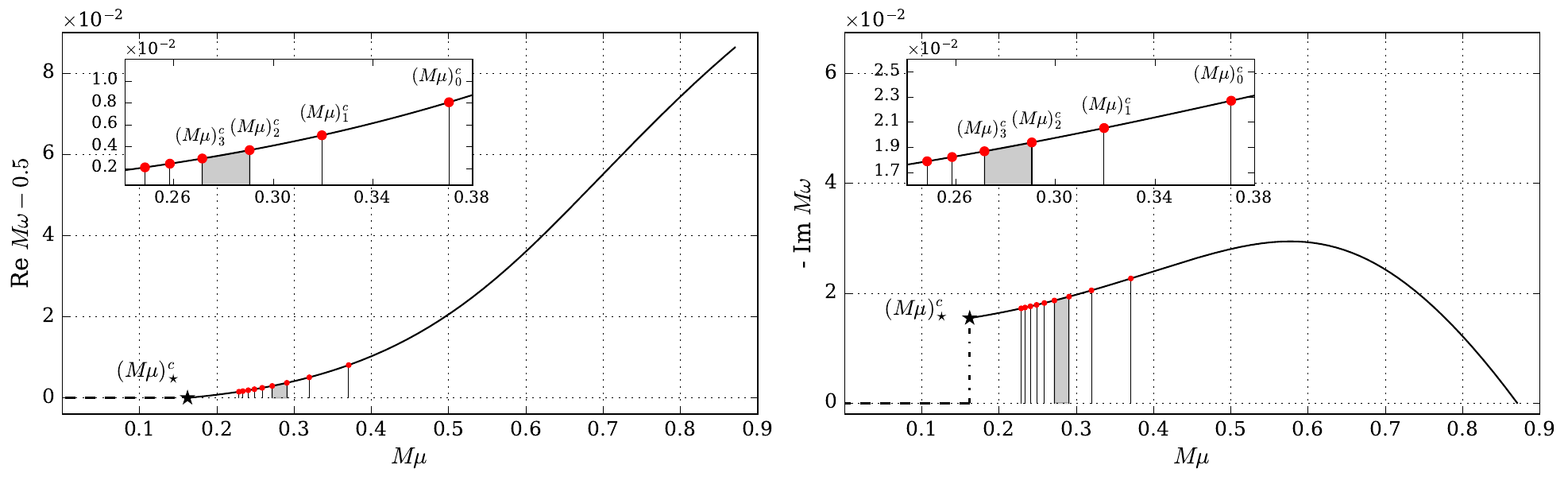}
	\caption{The real (left) and imaginary (right) parts of the DM frequency as functions of $M\mu$ in the extremal regime $a/M=1$ are shown by the solid curve originating at the star. The star marks the frequency associated with the lowest scalar mass for which this DM exists, namely $M\mu = (M\mu)^c_{\star}\simeq 0.16189$. Red dots denote the DM frequencies at $a/M=1$ and $M\mu = (M\mu)^c_n$. The shaded band corresponds to the interval $(M\mu)^c_3 < M\mu < (M\mu)^c_2$, which corresponds to region (b) of Fig.~\ref{fig:surfacebranching} through the projection of the $n=2$ and $n=3$ EPs onto the extremal Kerr line. Vertical lines illustrate how the spacing between EPs progressively narrows as one approaches the critical point $(M\mu)^c_{\star}\simeq 0.16189$, signaling the termination of the EP cascade. For visual clarity, we display only the first nine red dots.}
	\label{fig:criticalmassextremal}
\end{center}
\end{figure*}

When the scalar field is massless, all $(\ell,m)=(1,1)$ modes of the Kerr spectrum are ZDMs~\cite{PhysRevD.110.124064,PhysRevLett.133.261401}.  The spectrum of ZDMs in the extremal regime $a/M\rightarrow 1$, indexed by $k\in\mathbbold{N}$, is given by~\cite{PhysRevD.84.044046, PhysRevD.110.124064} 
\begin{equation}
\omega_{k} \simeq \frac{1}{2M} - i 2\pi  T_{+}\left(k+\frac{1}{2}\right) - i \pi  T_{+} \sqrt{4\lambda_0+4 M^2\mu^2-6},
\label{eq:omegaeqover}
\end{equation}
where $\lambda_{0} = (4 M^2 \mu^2+39)/20$ (see also~\cite{Hod:2008zz,Casals:2019vdb,daCunha:2021jkm,1980ApJ...239..292D,Cardoso:2004hh,Richartz:2017qep}). Since EPs mark transitions between ZDMs and DMs, we expect DMs to emerge in the spectrum if $M\mu$ is sufficiently large. In fact, we have confirmed that for sufficiently low masses $M\mu<(M\mu)^c_{\star}$, where $(M\mu)^c_{\star} \approx 0.16189$, the near extremal Kerr spectrum is comprised only by ZDMs for $(\ell,m)=(1,1)$. At $(M\mu)_\star^c$, we note the appearance of a single damped mode in the extremal spectrum, with frequency $(M\omega)^c_{\star} \approx 0.50000 - 0.015490 i $. We defer the technical details for the Supplemental Material~\cite{supplemental}, but, as explained below, we confirmed that it belongs to the spectrum by following it adiabatically in parameter space. By tracking this DM along the extremal line $a/M=1$, we determined that it remains part of the spectrum for at least $(M\mu)^c_{\star}  <  M\mu \le M\mu_{max}$. The value $M\mu_{max}\approx 0.8705$ reflects a limitation of the condition (S.18), as explained in \cite{PhysRevD.110.124064}, which is not relevant for the present discussion. The behavior of the DM at extremality ($a/M=1$) is shown in Fig.~\ref{fig:criticalmassextremal}.

\begin{figure}[htb!]
\begin{center}
	\includegraphics[width=0.48\textwidth]{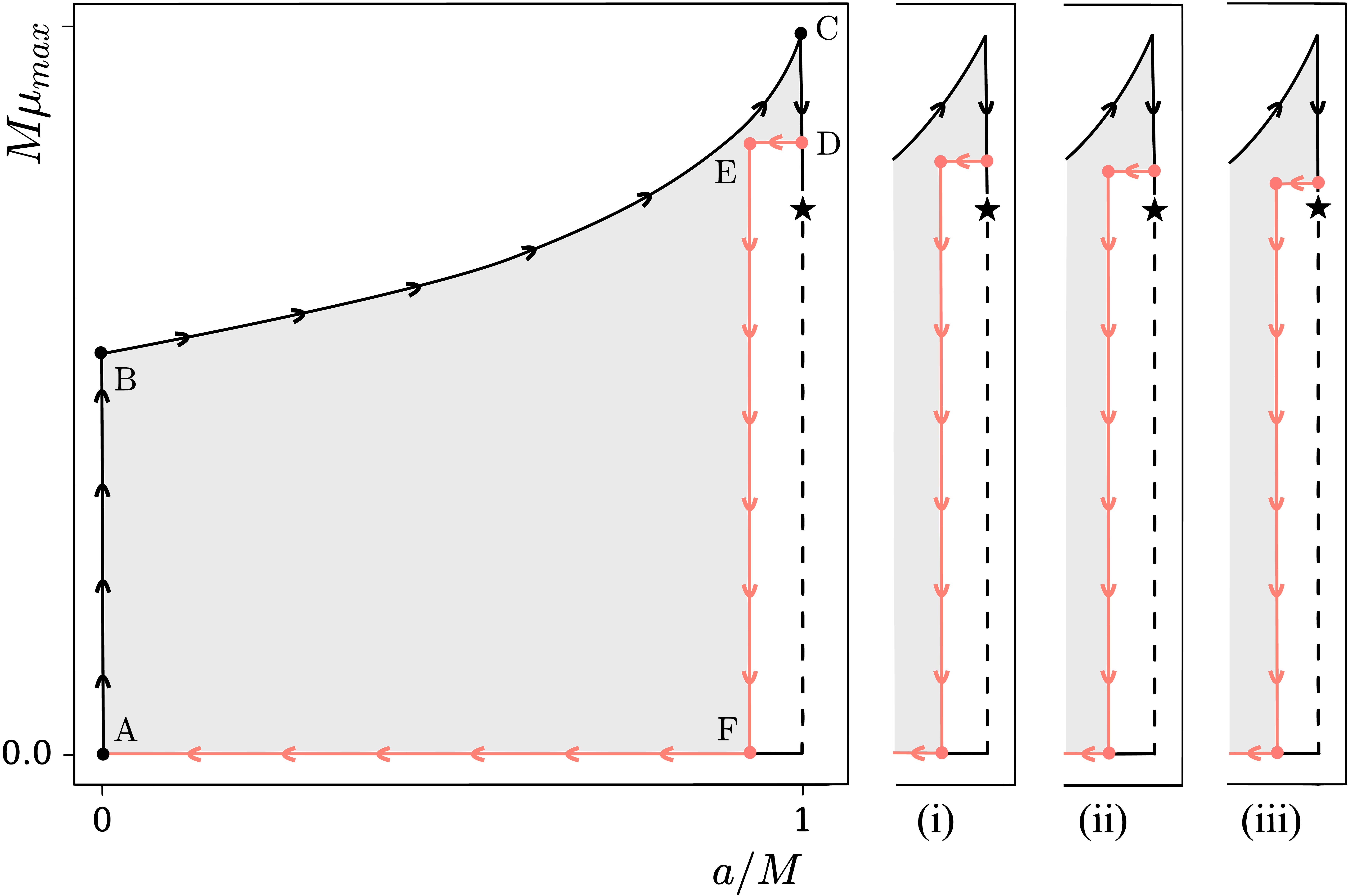}
\end{center}
\caption{The closed path ABCDEFA around the shaded region, beginning and ending at the origin, along which we adiabatically track the evolution of the Schwarzschild fundamental mode. The trajectories labeled (i), (ii), and (iii) illustrate different choices of $(M\mu)|_D=(M\mu)|_E$, each enclosing a different number of EPs within the shaded region.} 
\label{fig:ParSpal2m2}
\end{figure}

We follow the DM adiabatically along a path that starts at the line $a/M=1$ and ends at the origin $(a/M, M\mu) = (0,0)$. Because EPs are present, the outcome depends not only on the choice of the starting point but also on the path itself. For definiteness, we adopt the trajectory DEFA shown in Fig.~\ref{fig:ParSpal2m2}. We first lower the spin from $a/M=1$ to $a/M=0.99$ while keeping $M\mu=(M\mu)|_D$ fixed. Next, we decrease the mass $M\mu$ to zero, at fixed $a/M$. Finally, we reduce $a/M$ to zero, keeping $M\mu=0$, and thus reach the origin. The resulting mode is then compared against the Schwarzschild overtone spectrum, allowing us to associate each starting value $(M\mu)|_D$ with the appropriate overtone number $n$.

We find that this procedure yields arbitrarily large values of $n$ as $(M\mu)|_D$ approaches (from above) the critical mass $(M\mu)^c_{\star}$. This is another evidence for the presence of an infinite sequence of EPs, following the cascade shown in Fig.~\ref{fig:ImReomega}, which accumulate at the point $(a/M,M\mu)=(1,(M\mu)^c_{\star})$. In contrast, increasing the initial value of $(M\mu)|_D$ decreases the overtone number $n$ obtained at the end of the procedure. For example, if the starting mass lies in the interval $(M\mu)^c_3 \le (M\mu)|_D < (M\mu)^c_2$ (see the shaded band in Fig.~\ref{fig:criticalmassextremal}), the procedure yields $n=3$. If the initial mass satisfies $(M\mu)|_D \ge (M\mu)^c_0$, it returns the fundamental mode $n=0$. Hence, we conclude that the geometric phases associated with EPs are ergodic:
they enable access to every overtone index $n$ in
the massless Schwarzschild limit. More precisely, by choosing the
initial scalar mass at extremal spin within the interval $(M\mu)^c_n
\le (M\mu)|_D < (M\mu)^c_{n-1}$, the corresponding mode evolves into
the $n$-th overtone, meaning that all overtones are reachable
through appropriate adiabatic paths.

\noindent \textbf{\textit{Ergodic hysteresis and EP counting   -- }} Evidence for an infinite EP cascade, together with the ergodicity of the associated geometric phases, leads to the notion of \textit{ergodic hysteresis}: any state at a fixed point in the parameter space $(a/M, M\mu)$ can be adiabatically connected to any other state at the same point, provided the chosen path encloses a suitable combination of EPs. For example, starting from the fundamental QNM at $(a/M,M\mu) = (0,0)$, if one adiabatically follows a path that encloses the first $n$ EPs and returns to the origin, the system ends up in the $n$th overtone rather than the fundamental mode. Even if the precise locations of the EPs are unknown, this mechanism offers a way to determine how many of them lie inside a given closed path. 

To illustrate ergodic hysteresis and the EP counting strategy, we use the case $(\ell,m)=(2,2)$ as an example.  Starting at the origin of the parameter space, we track how the fundamental mode changes as we follow the closed contour ABCDEFA around the shaded region shown in Fig.~\ref{fig:ParSpal2m2}. This region is bounded from above by the curve BC along which the fundamental mode has $\mathrm{Im}(M\omega) = 0$, extending from $(a/M,M\mu) \simeq (0,0.821)$ to $(1,1.476)$. Along the extremal line $a/M=1$, the spectrum exhibits a single DM for $1.114 \lesssim M\mu \lesssim 1.476$.  We fix $(a/M)|_E = (a/M)|_F = 0.99$ and vary $(M\mu)|_D = (M\mu)|_E$, so the trajectory DEF determines how many EPs are enclosed by the loop. For instance, when $(M\mu)|_D = 1.170$, the mode obtained after completing the loop no longer corresponds to the fundamental mode, but instead to the first overtone. In this case, the trajectory encloses a single EP where these two modes coalesce. We estimate the location of this degeneracy to be approximately $(a/M,M\mu) \simeq (0.9999,1.1717)$. More generally, as shown in Table \ref{tab:l2m2_paths}, different choices of $(M\mu)|_D$ result in different overtones after completing the loop, allowing us to infer how many EPs are enclosed. For example, choosing $(M\mu)|_D = 1.1200$ yields the 26-th overtone, implying 26 EPs within the path. As $(M\mu)|_D$ approaches the critical value $(M\mu)^{c}_{\star} \simeq 1.114$, the overtone number increases significantly ($n \gg 1$), enabling the inference of $n$ (and thus the number of EPs) through the asymptotic formula of \cite{Motl:2002hd} (or its generalizations~\cite{Neitzke:2003mz,MaassenvandenBrink:2003as,Musiri:2003bv}).

\begin{table}[hbt!]
\begin{ruledtabular}
	\begin{tabular}{cccccc}
		$(M \mu)|_D$ & $n$ & $M\omega_n$  & $(M \mu)|_D$ & $n$ & $M\omega_n$ \\ \hline
		1.1450 & 2 & 0.43054 - 0.50855i & 1.1275 & 9  & 0.27146 - 2.23933i\\
		1.1400 & 3 & 0.39386 - 0.73809i & 1.1250 & 12 & 0.24547 - 2.99854i\\
		1.1350 & 5 & 0.33489 - 1.22841i & 1.1225 & 16 & 0.22258 - 4.00841i\\
		1.1300 & 7 & 0.29697 - 1.73276i & 1.1200 & 26 & 0.18979 - 6.52486i\\
	\end{tabular}
\end{ruledtabular}
\caption{Mapping of the fundamental QNM to the $n$th overtone after one loop around the path ABCDEFA in Fig.~\ref{fig:ParSpal2m2}, which begins and ends at the origin $(a/M,M\mu)=(0,0)$. The value of $(M\mu)|_D=(M\mu)|_E$ determines the final frequency and, therefore, the number of EPs enclosed by the path.}
\label{tab:l2m2_paths}
\end{table}

\noindent \textbf{\textit{Discussion --}} We presented evidence above
for infinite EP cascades in the $(\ell,m)=(1,1)$ and $(2,2)$ sectors
of the Kerr QNM spectrum of massive scalar perturbations, arising in
the near-extremal regime through ZDM–DM transitions induced by
accidental degeneracies. The associated geometric phases mix in such a
way that, by choosing an appropriate path in parameter space, any two
QNMs at arbitrary points can be connected
adiabatically.
This allows adiabatic variations to arbitrarily reorganize the spectrum: by choosing an appropriate deformation path, the spectral ordering can be reshuffled into any other ordering. We refer to this phenomenon as \emph{adiabatic ergodicity}. Such path-dependent reorganization undermines the usual meaning of quantum numbers: the overtone label is no longer determined solely by the spectral ordering at a given point in parameter space, but also by the adiabatic path connecting the initial and final configurations.
Given the structure revealed for $(\ell,m)=(1,1)$ and $(2,2)$, we expect EP cascades and adiabatic ergodicity to arise throughout the entire $m \ge 1$ spectrum. Our findings reinforce the emerging view of rotating black holes as natural laboratories for non-Hermitian physics and EP dynamics. 

In particular, the existence of an infinite number of EPs raises the intriguing possibility of EP-based mode control in black hole dynamics. During a ringdown phase dominated by the lowest overtones, adiabatic evolution of the system parameters along a path enclosing several points of an infinite EP cascade could, in principle, redistribute energy among overtones in a controlled manner, exciting higher modes while de-exciting lower ones. Whether such behavior can occur dynamically in black holes is an open question. Notably, in photonics and optomechanics, mode and energy manipulation through EP encirclement is already an active area of research~\cite{Doppler:2016pek,PhysRevLett.118.093002,Zhaoscience,PhysRevLett.124.153903,Patil:2021jmy,Long:2022dyt,Guria:2023eqk}, suggesting nontrivial gravitational analogs yet to be explored.

\noindent \textbf{\textit{Data Availability --}} Data supporting the findings of this work are openly available~\cite{cavalcante_2025_15619940}.

\noindent \textbf{\textit{Acknowledgements --}} The authors thank J. Barragán-Amado for comments and suggestions on the manuscript. J.~P.~C. acknowledges the financial support from the S\~ao Paulo Research Foundation (FAPESP, Brazil), Process Number 2024/15921-9. M.~R.~acknowledges partial support from the Conselho Nacional de Desenvolvimento Cient\'{i}fico e Tecnol\'{o}gico (CNPq, Brazil), Grant 315991/2023-2, and from the S\~ao Paulo Research Foundation (FAPESP, Brazil), Grants 2022/08335-0 and 2024/00923-6.

\section{Supplemental Material}

\noindent \textbf{\textit{The spectrum near EPs --}} The Riemann-Hilbert map \eqref{eq:rhmap}, along with the Fredholm determinant formulation of the Painlevé tau functions, casts the equations satisfied by the QNMs into a type of non-linear secular equation. This was pointed out in \cite{PhysRevD.110.124064}, and the procedure is as follows. The accessory parameter $c_t$ has an expansion in terms of the modulus $t$:
\begin{equation}
  c_t=c_{-1}(\sigma)t^{-1}+c_0(\sigma)+c_{1}(\sigma)t+\ldots,
\end{equation}
where we have omitted the dependence of the coefficients $c_k$ on the single monodromy parameters $\{\theta_k\}=\{\theta_1,\theta_2,\theta_\star\}$. The explicit formula can be obtained, for instance, using the continued fraction method of \cite{daCunha:2022ewy}. This expansion can be formally inverted as
\begin{equation} \label{sig_exp}
  \sigma = \sigma_0+\sigma_1t+\sigma_2t^2+\ldots,
\end{equation}
with the coefficients $\sigma_k$ given by rational functions of $\theta_1$ and $\theta_2$, and polynomials in $\theta_\star$ and $c_t$. The first two terms are
\begin{equation}
  \sigma_0=\theta_1+\theta_2,\quad
  \sigma_1=\frac{2(\theta_1+\theta_2-2)c_t-\theta_\star(\theta_2-1)}{
    (\theta_1+\theta_2-1)(\theta_1+\theta_2-2)}.
\end{equation}

After inserting \eqref{sig_exp} into \eqref{eq:quantcond}, we can define $\eta$ as a function of the parameters of the CHE, namely $\{\theta_k\}$, $t$ and $c_t$. One also needs to solve the zero condition of the tau function, i.e.~equation \eqref{eq:rhmap:a}, which, up to an elementary function, can be written in terms of a determinant of an integral operator $\mathsf{K}$~\cite{Lisovyy:2018mnj,CarneirodaCunha:2019tia}:
\begin{equation}
  \det(\mathbbold{1}-\mathsf{K}(\{\theta_k\};c_t,t))=0.
  \label{eq:fredholmPV}
\end{equation}
For the sake of the argument, we assume that $\mathsf{K}$ has finite rank and can thus be represented as a matrix in an appropriate basis. In our application, the parameters of the equation above are functions of $M\omega$, $a/M$ and $M\mu$, for fixed quantum numbers $\ell,m$ (we assume that the angular eigenvalue problem has been solved). We thus arrive at a working definition $\mathsf{K}=\mathsf{K}(M\omega,a/M,M\mu)$. 

At a generic point in the parameter space, the determinant has isolated zeros in $M\omega$, which correspond to QNM frequencies. A given change of $a/M$ or $M\mu$ produces a change in the QNM frequency. Linearization of \eqref{eq:fredholmPV} implies that the variations $\delta(M\omega)$, $\delta(a/M)$, and $\delta(M\mu)$ are related by
\begin{equation}
  K_{M\omega}\delta(M\omega)+K_{a/M}\delta(a/M)+K_{M\mu}\delta(M\mu)=0,
  \label{eq:linearization}
\end{equation}
where
\begin{equation}
  K_{X}=\Tr\left[\adj(\mathbbold{1}-\mathsf{K})
    \frac{\partial \mathsf{K}}{\partial X}
  \right],
\end{equation}
with $X=\{M\omega,a/M,M\mu\}$. The adjugate $\adj \mathsf{A}$ of a matrix $\mathsf{A}$ is the transpose of the matrix of cofactors and, for invertible matrices, is related to the inverse by $\adj(\mathsf{A})=\det(\mathsf{A})\,\mathsf{A}^{-1}$. Unlike the inverse, however, the adjugate is well-defined even when $\det\mathsf{A}=0$.

We assume that the existence of an EP is due to a double zero of \eqref{eq:fredholmPV}, meaning that the associated QNM frequecies are doubly degenerate. In this case, the linearization \eqref{eq:linearization} is not enough to describe infinitesimal changes, and we have to consider higher order terms in the expansion. This means that, at an EP, the following condition is satisfied.
\begin{equation}
  K_{M\omega}=\Tr\left[\adj(\mathbbold{1}-\mathsf{K})
    \frac{\partial \mathsf{K}}{\partial (M\omega)}
  \right]=0.
\end{equation}
In addition, near an EP, the linearization \eqref{eq:linearization} must be replaced by 
\begin{equation}
  K'_{M\omega}(\delta(M\omega))^2+K_{a/M}\delta(a/M)+K_{M\mu}\delta(M\mu)=0,
  \label{eq:linearization2}
\end{equation}
where 
\begin{multline}
  K'_{M\omega}=-\frac{1}{2}\Tr\left[\adj(\mathbbold{1}-\mathsf{K})
    \frac{\partial \mathsf{K}}{\partial (M\omega)}\right]^2\\
  -\frac{1}{2}\Tr\left[\adj(\mathbbold{1}-\mathsf{K})
    \frac{\partial^2 \mathsf{K}}{\partial (M\omega)^2}
  \right].
\end{multline}
Note that \eqref{eq:linearization2} is equivalent to \eqref{eq:nearEPbehavior}. In particular, the coefficients $A_n$ and $B_n$ of \eqref{eq:nearEPbehavior} take the form 
\begin{equation} \label{AnBn}
  A_n=\frac{K_{a/M}}{K'_{M\omega}},\qquad 
  B_n=\frac{K_{M\mu}}{K'_{M\omega}},
\end{equation}
where the expressions on the right hand side must be evaluated at the EP indexed by the integer $n$, i.e.~($(M\omega)^c_n,(a/M)^c_n,(M\mu)^c_n$). These quantities can, in principle, be computed by truncating $\mathsf{K}$ and using the Cayley-Hamilton theorem to express the adjugate in terms of a polynomial in $\mathsf{K}$, but the process is computationally demanding. Instead, we fit \eqref{eq:nearEPbehavior} to the numerical data, as shown in Table \ref{tab:AnBnl1m1}.

\begin{figure}[htb!]
\begin{center}
\includegraphics[width=0.48\textwidth]{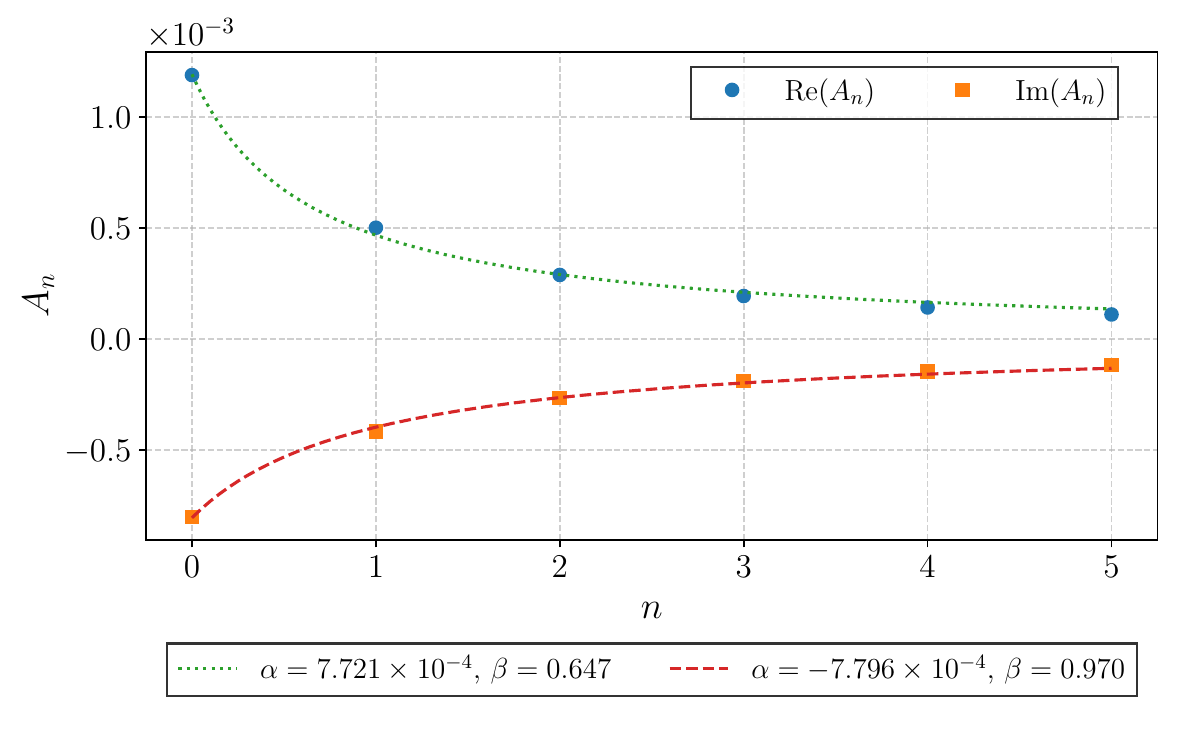}
\end{center}
\caption{The real (circle dots) and imaginary (square dots) parts of the  coefficient $A_n$ as a function of $n$, as given in Table \ref{tab:AnBnl1m1} -- see also Eqs.~\eqref{eq:nearEPbehavior} and \eqref{AnBn}. The dashed and dotted  lines correspond, respectively, to fits of $\mathrm{Re}(A_n)$ and $\mathrm{Im}(A_n)$ with respect to an harmonic sequence of the form $\alpha/(n + \beta)$.}
\label{fig:ImReAn} 
\end{figure}

The behavior of the coefficients $A_n$ is particularly interesting. In Fig.~\ref{fig:ImReAn}, we show the real and imaginary parts of $A_n$ as a function of $n$. Both series of data in Table \ref{tab:AnBnl1m1} are well fitted by harmonic sequencies of the form $\alpha/(n + \beta)$, with $(\alpha,\beta)=(7.721 \times 10^{-4},0.647)$ fitting the data for $\mathrm{Re}(A_n)$ and $(\alpha,\beta)=(-7.796 \times 10^{-4},0.970)$ fitting the data for $\mathrm{Im}(A_n)$. In particular, these fits suggest that there is an infinite number of EPs, and that, in the limit $n \rightarrow \infty$, the coefficient $A_n$ associated with the $n$-th EP tends to zero.

\noindent \textbf{\textit{Extremal Kerr and the $\tau_{\tiny III}$-function --}} \label{sec:Conflimit_and_Kerrextremal} 
For given values of the angular quantum numbers $\ell,m>0$, the imaginary part of a QNM frequency may or may not vanish at the extremal point $a/M=1$, characterizing a ZDM or a DM, respectively. In terms of the parameters of the CHE \eqref{eq:che}, extremality corresponds to the confluent limit:
\begin{equation}
\begin{aligned}
\Lambda = \frac{1}{2}(\theta_2+\theta_1), \ \ \ \theta_{\circ}
  =\theta_2-\theta_1, \ \ \ x = \Lambda z
, \ \ \ u = \Lambda t,
\end{aligned}
\label{eq:conflimit}
\end{equation}
with $\Lambda \rightarrow \infty$. This brings \eqref{eq:che} to the form
\begin{equation}
\frac{d^2
  h}{dx^2}+\bigg[\frac{2-\theta_{\circ}}{x}-\frac{u}{x^2}\bigg]
\frac{dh}{dx}-\bigg[\frac{1}{4} 
+\frac{\theta_{\star}}{2x}+\frac{uc_{u}-u/2}{x^2}\bigg]h=0, 
\label{doubheuneq}
\end{equation}
with $uc_u = \lim_{\Lambda \rightarrow\infty}tc_t$. The equation above is the double-confluent Heun equation (DCHE), an ordinary differential equation with two irregular singularities of Poincaré rank 1 at $z=0$ and $z=\infty$~\cite{NIST:DLMFc13}. 

We have to solve simultaneously the angular and radial eigenvalue problems to determine QNMs. The angular problem has a smooth extremal limit and, hence, even in the case of extremal black holes can be formulated using the Painlevé V transcendent~\eqref{eq:rhmap}~\cite{CarneirodaCunha:2019tia}. In particular, the angular eigenvalue $ \lambda_{\ell,m}$ can be expanded as
\begin{multline}
  \lambda_{\ell,m} =\ell (\ell+1)-\frac{\left(2 \ell(\ell+1)-2 m^2-1\right) }{
    4\ell(\ell+1)-3}M^2\left(\omega^2-\mu ^2\right)+\\
  +\mathcal{O}(M^4(\omega^2-\mu^2)^2).
\end{multline}
The analysis of the radial equation, on the other hand, is more involved. The parameter $\Lambda$ in this case can be expanded, near extremality, as
\begin{equation}
\Lambda =\frac{i(2 M\omega-m)}{\delta }+\frac{1}{3} i \delta  (m+M\omega)+\mathcal{O}(\delta^2),
\label{eq:conflimitLamb}
\end{equation}
with $\delta$ defined through $a/M=\cos\delta$ (when $\delta=0$, $a/M=1$ and the black hole is extremal). We then see that $\Lambda$ diverges in the extremal limit unless $M\omega\rightarrow m/2$, which is the case for DMs. 
The corresponding parameters for the radial DCHE are
\begin{equation}
\begin{aligned}
\theta_{\circ} =4 iM\omega,\qquad
\theta_{\star} = i\frac{2M(2 \omega^2-\mu^2)}{ \sqrt{\omega^2-\mu ^2}}.
\end{aligned}
\label{eq:extremalparameters}
\end{equation}
Moreover, the accessory parameter $c_{u}$ and the modulus $u$ are:
\begin{subequations}
\begin{gather}
  u= 4M(m-2 M\omega)\sqrt{\omega^2-\mu^2},\\
  u\,c_{u} =\lambda_{l,m}+\tfrac{1}{2}u
  +M^2(\mu^2-7\omega^2)-\tfrac{1}{4}(\theta_\star-2)\theta_\star.
\end{gather}
\end{subequations}

As discussed in \cite{Cavalcante:2021scq}, the Riemann-Hilbert map for the DCHE is written in terms of the tau function for the Painlevé III transcendent as
\begin{equation}
\begin{gathered}
\tau_{\tiny III}(\theta_\star,\theta_\circ;\sigma,\eta;u)=0, \\
\frac{d}{du}\text{log}
\tau_{\tiny III}(\theta_\star-1,\theta_\circ-1;\sigma-1,\eta;u)-
\frac{(\theta_{\circ}-1)^2}{2u}=
c_{u},
\end{gathered}
\label{eq:rhmapIII}
\end{equation}
where $\sigma$ and $\eta$ are monodromy parameters that play essentially the same role as their non-confluent counterparts when $0\leq a/M < 1$. Small $u$ expansions for $\eta$ and $c_u$ as functions of $\theta_\star,\theta_\circ$, and $\sigma$ are available in \cite{Cavalcante:2021scq}. For a more comprehensive review of the mathematical framework of \eqref{eq:rhmapIII}, we also refer the reader to \cite{cavalcante2023isomonodromy}.   

We assume $\mathrm{Re}(M\omega)  > m/2$, ensuring that $\Lambda$ diverges in the upper half-plane as $\delta\to 0$. Since $\mathrm{Im}(M\omega)<0$, this divergence occurs in the first quadrant. Under this assumption, the QNM condition \eqref{eq:quantcond} in the extremal limit reduces to
\begin{equation}
e^{i\pi\eta}=e^{-2\pi i\sigma}
\frac{\sin\tfrac{\pi}{2}(\theta_\star+\sigma)}{
	\sin\tfrac{\pi}{2}(\theta_\star-\sigma)}
\frac{\sin\tfrac{\pi}{2}(\theta_\circ+\sigma)}{
	\sin\tfrac{\pi}{2}(\theta_\circ-\sigma)}.
\label{eq:quantizationIII}
\end{equation}
Using the Riemann-Hilbert map \eqref{eq:rhmapIII} and the quantization condition \eqref{eq:quantizationIII}, we determine the DMs associated with the extremal spin $a/M=1$.
In the restricted domain $\mathrm{Re}(M\omega) >  m/2$, and focusing on the cases $(\ell,m)=(1,1)$ and $(2,2)$, we find that the correct QNM flux conditions at both the outer horizon and infinity are satisfied only for $(M\mu)_\star^c <  M\mu \leq M\mu_{\max}$. Within this mass range, the resulting solutions are DMs. As a further consistency check, we then followed these DMs adiabatically for $a/M<1$, down to the origin of the parameter space, and verified that they coincide with modes in the Schwarzschild overtone spectrum.

We have also investigated the existence of modes with Re$(M\omega)<m/2$. By tracking the corresponding solutions throughout the parameter space, we found that they fail to reproduce known QNMs in the limit $a/M \rightarrow 0$. A closer examination of these spurious solutions shows that they do not satisfy the standard QNM flux conditions at either the outer horizon or at infinity.

\bibliography{prl_2025.bib}

\end{document}